\documentclass[preprint,aps,prl]{revtex4-2}
\usepackage{graphicx}
\usepackage{textcomp}
\usepackage{amsmath}
\usepackage{fontenc}
\usepackage{subfigure}
\usepackage{tabularx}
\usepackage{float}
\usepackage{xcolor}
\usepackage[english]{babel}
\usepackage[active]{srcltx}

\begin{document}
\title{A Variational Description of the Adsorption Phenomenon.}
\author{Giovanni Barbero$^{1,2}$, Anca Luiza Alexe Ionescu $^{1,3}$, Ioannis Lelidis $^{4}$}
\affiliation{$^1$Dipartimento di Scienza Applicata e Tecnologia, Politecnico di Torino, Corso Duca degli Abruzzi 24, Torino 10129, Italy.\\
$^2$Istituto dei Sistemi Complessi (ISC--CNR), Via dei Taurini 19, 00185 Rome, Italy.\\
$^3$University Politehnica of Bucharest, Faculty of Applied Sciences, Splaiul Independentei 313, 060042, Bucharest.\\
$^4$Department of Condensed Matter Physics, Faculty of Physics, National and Kapodistrian University of Athens, Panepistimiopolis, Zografos, Athens 157 84, Greece}

\date{\today}
\begin{abstract}
\section{Abstract}
The adsorption of particles dispersed in a liquid is analyzed introducing a variational approach. While traditional adsorption models rely on localized, discrete site-binding kinetics, we treat the system as a binary mixture consisting of adsorbate particles and the host liquid within a continuous framework. An expression for the variation of the mixing entropy arising from an inhomogeneous distribution of the adsorbate, of density $n(z)$, is derived. The interactions between adsorbate particles are modeled by a bulk free energy functional containing a term proportional to the square of the density gradient, $\nabla n$, and a bulk compressibility term which exhibits a well-defined minimum at the equilibrium bulk density of adsorbate particles, $n_0$, in the absence of adsorption.  This approach offers the distinct advantage of capturing non-local fluid structures and particle-particle correlations near interfaces that discrete models overlook. The profile $n(z)$ is obtained analytically by linearization of the Euler-Lagrange equilibrium equation for a finite sample confined between two parallel walls under the constraint of particles number conservation. The surface density of adsorbed particles, $\sigma$, is then calculated for the simplified case in which the interaction with the substrate is short-ranged and decays exponentially into the bulk. For a semi-infinite sample limited by a planar wall, the analysis predicts that $\sigma \propto n_0$ for small values of $n_0$, whereas it saturates at a constant value for large $n_0$, in agreement with the predictions of the classical Langmuir model, thereby providing a thermodynamic foundation for phenomenological adsorption isotherms. The case in which, in addition to the standard adsorption energy, a localized surface energy is present, is also considered. In this situation, an effective adsorption energy associated with the localized contribution is derived in the limiting cases of vanishing and very large anchoring strength, offering a predictive tool for tailoring surface modifications in experimental soft matter systems.

\end{abstract}
\maketitle
\section{I. Introduction}

The adsorption phenomenon is related to the existence of surface forces between the limiting surface and the particles dissolved in the liquid in contact with it \cite{israelachvili}. This phenomenon is usually described by means of Langmuir model, which is valid in the limit of small adsorption \cite{langmuir}. In this model, the kinetic equation describing the adsorption rate contains two terms: one associated with adsorption and the other with desorption. These two terms are related to two phenomenological coefficients known as the adsorption coefficient and the desorption time. The relation  between these phenomenological parameters and the interaction energy between the molecules and the substrate has been discussed in \cite{luiz}.

However, classical discrete models like the Langmuir or Fowler-Guggenheim isotherms \cite{langmuir,ioannis} treat the adsorbent surface as a collection of isolated sites and often neglect the continuous structural variations of the fluid medium adjacent to the wall. In complex soft matter systems such as colloidal suspensions, nanoparticle dispersions, or polymer solutions, the localized density variations and long-range steric or electrostatic interactions create an inhomogeneous interphase zone. While modern Density Functional Theory and molecular dynamics simulations can capture these microscopic details, they are often computationally expensive and lack closed-form analytical clarity. Therefore, there is a clear theoretical gap for a macroscopically accessible continuum thermodynamic framework that can reconcile nonlocal particle-particle correlations with macroscopic adsorption isotherms.

In the present paper, we propose an approach to the adsorption phenomenon based on a continuum description of the medium in which the adsorbate particles are distributed. Assuming that, in a unlimited sample, the bulk density of adsorbate particles at thermodynamic equilibrium has a well defined value, we build an expression for the free energy density in the presence of an adsorbing surface responsible for a variation of the bulk density of the adsorbate particles. Besides the direct interaction between the adsorbate particles and the substrate \cite{israelachvili,liquids}, the free energy density includes a compression energy related to the local deviation of the density from its equilibrium value, as well as a gradient term associated with the spatial variation of the density itself. It also contains an entropic contribution originating from deviations of the bulk density from the equilibrium value.
The analysis is carried out by minimizing the total free energy of the sample, both in the absence and in the presence of a localized surface energy. Conservation of the number of particles is enforced by means of a Lagrange multiplier, following a standard variational procedure. The simple case in which the sample occupies a half-space is also considered.

The primary advantage of the proposed variational framework is its ability to analytically capture the smooth transition of the fluid's microstructure near a confining wall via the gradient term ($\nabla n$), while recovering the classic Langmuir behavior ($\sigma \propto n_0$ at low concentrations and saturation at high concentrations) as a natural thermodynamic consequence rather than an a priori kinetic assumption. Furthermore, by incorporating a localized surface energy, our model introduces a versatile method to evaluate anchoring physics, which is highly relevant to liquid crystal alignments and polymer wetting.

From an experimental standpoint, this continuum model is particularly well-suited to interpret adsorption profiles in systems where the adsorbate size is comparable to the characteristic interaction length of the substrate. For instance, the exponential decay of the substrate potential assumed herein closely mirrors the screened electrostatic double-layer interactions in dilute electrolytes or the steric depletion forces in polymer-colloid mixtures. By measuring the surface density $\sigma$ via optical techniques such as Ellipsometry or Surface Plasmon Resonance (SPR) as a function of bulk concentration, experimentalists can utilize our analytical expressions to directly extract the bulk compressibility and elastic properties of the host medium without relying on empirical fitting parameters.

The paper is organized as follows. In Sect.II a few phenomenological models are critically analyzed and their limits and merits discussed. In Sect. III, the variation of the mixing entropy term entering the free energy density is briefly discussed, and  the bulk energy density describing the interaction between the particles is presented  where two phenomenological parameters related to the compressibility and elastic properties of the medium are introduced. In Sect. IV, we present the variational problem  and the solution of the corresponding Euler–Lagrange equation. The presence of a localized surface energy is considered in the same Section. The conclusions are presented in Sect. V.
In Appendix 1, a justification of the energy density used in the analysis is provided, whereas Appendix 2 discusses in detail the case in which the sample occupies a half-space.

\section{II. Phenomenological models}
In this section, we present a few phenomenological models for adsorption that, in some situations, are equivalent to Langmuir's model. Versions of these models have been used to fit experimental data by several researchers \cite{x1,x2,x3,x4}. The main advantage of these models, as well as of their generalizations discussed below, are their simplicity and the reduced number of phenomenological parameters appearing in the formulas for the surface density of adsorbed particles on the adsorbing surface. All of these models are based on the assumption that the bulk density of adsorbable particles is uniform across the sample, as a consequence of the short range of the surface forces responsible for adsorption. They completely neglect what happens within the surface layer, whose thickness is comparable to the range of the surface forces. Instead, they introduce the concept of surface density, which is related to the variation of the bulk density of adsorbable particles caused by the presence of the adsorbing surface. These models are phenomenological, i.e., they are based on reasonable assumptions concerning adsorption and desorption processes and their dependence on the density of particles already adsorbed and those located just in front of the adsorbing surface.

Let us consider a sample in the shape of a slab of thickness $d$. The liquid filling the sample has, in thermodynamic equilibrium (i.e. in an unlimited sample), a bulk density of adsorbable particles $n_0$. We assume that the adsorbed particles can be described by a surface density of particles $\sigma$. The kinetic equation relating $\sigma$ to the bulk density of particles just in front of the adsorbing surface, $n_s$, is
\begin{equation}
\label{x1}\frac{d\sigma}{dt}={\cal A}[n_s(t),\sigma(t)]-{\cal D}[n_s(t),\sigma(t)],
\end{equation}
where ${\cal A}[n_s(t),\sigma(   t)]$ and ${\cal D}[n_s(t),\sigma(t)]$ represent the adsorbing and the desorption terms, respectively. Equation (\ref{x1}) is valid in the case in which the surface forces responsible for the adsorption are short-range,such that it is possible to define $\sigma$ as the integral of the bulk variation of the density of particles over the range of the surface forces. Of course, the presence of the surface forces is responsible for a spatial dependence of the bulk density of adsorbable particles. In the absence of adsorbing surfaces, the bulk density is $n=n_0$, and in the presence of the adsorbing surfaces $n=n(z)$, where the $z$-axis is normal to the limiting surfaces, placed at $z=\pm d/2$ and assumed identical in our analysis. Due to the conservation of the number of adsorbable particles, we have the condition
\begin{equation}
\label{x2}2 \sigma(t)+\int_{-d/2}^{d/2} n(z,t)\,dt=n_0 d.
\end{equation}
Equation (\ref{x1}) indicates that the adsorption phenomenon is a dynamical effect \cite{y1}. To solve the problem and find $\sigma=\sigma(t)$ and the profile of the bulk density of the adsorbable particles $n=n(z,t)$, it is necessary to relate $n(z,t)$ to the potential energy giving rise to the adsorption force.

At equilibrium, the adsorption and desorption contributions are equal for all $t$, that is ${\cal A}[n_s(t),\sigma(t)]={\cal D}[n_s(t),\sigma(t)]$,
and $\sigma$ as well as $n_s$ have time independent values.

The functional form of ${\cal A}[n_s(t),\sigma(t)]$ and ${\cal D}[n_s(t),\sigma(t)]$ can be constructed taking into account that the adsorption term is expected to increase with increasing $n_s$ (the density of adsorbable particles just in front of the adsorbing surface), and the desorption term is expected to increase with $\sigma$ (the density of particles already adsorbed). Of course, the experimental data will validate the proposed functional forms.
\subsection{II.1 Without Surface Saturation}
In a very simple approach, valid only in the limit of small $\sigma$ compared to the maximum surface density of adsorption $\sigma_0$, it is reasonable to put
\begin{equation}
\label{x4}{\cal A}[n_s(t),\sigma(t)]=\kappa n_s(t),\quad {\cal D}[n_s(t),\sigma(t)]=\frac{1}{\tau} \sigma(t),
\end{equation}
where $\kappa$ and $\tau$ are the adsorption coefficient and desorption time, measured in $m/s$ and $s$, respectively. In this framework kinetic equation (\ref{x1}) takes the form
\begin{equation}
\label{x5}\frac{d\sigma}{dt}=\kappa\, n_s(t)-\frac{1}{\tau}\,\sigma(t).
\end{equation}
This equation has been used to describe the electric response of a medium to an external periodic excitation of small amplitude in the framework of standard \cite{z1,z2,z3,z4} and fractional \cite{w1,w2,w3} time derivatives, respectively.

Assuming that the surface forces responsible for the adsorption are short-range, in the bulk $n(z,t)$ is position independent, $n(z,t)=n(t)$ and differs from the equilibrium value $n_0$. In this framework, Eq.(\ref{x2}), representing the conservation of particles, can be written as
\begin{equation}
\label{x2-1} 2 \sigma(t)+n_s(t)\,d=n_0 d.
\end{equation}

At equilibrium, from (\ref{x5},\ref{x2-1}) we get
\begin{eqnarray}
\label{x8}n_s=n_0\,\,\frac{1}{1+2(\kappa \tau/d)}\quad\&\quad\sigma=n_s \kappa \tau\,.
\end{eqnarray}
In the framework of this model, $\kappa \tau$ is a characteristic length associated with the adsorption phenomenon \cite{langmuir}.
For $d\ll \kappa \tau$,
$\sigma\to n_0\,d/2$ and  $n_s\to \sigma/(\kappa \tau)$,
whereas for $d\gg \kappa \tau$, $n_s\to n_0$ and $\sigma\to n_0\,\kappa \tau$.

According to the model based on (\ref{x4},\ref{x5}), the quantity $\sigma$ is proportional to $n_0$. This means that $\sigma$ diverges for $n_0\to \infty$, which is an unphysical result since $\sigma$ has to tend to a saturation value in concentrated solutions. It follows that the present model works well only in the limit of small $n_0$. Assuming a characteristic size $R$ for the adsorbed particle, the maximum surface density of adsorbed particles is of the order
$\sigma_0=1/(\pi R^2)$, and the model can correctly describe adsorption only for $\kappa \tau \,n_0\ll \sigma_0$
i.e., for $n_0\ll 1/(\pi R^2 \kappa \tau)$.

The case considered above concerns a closed system, in the sense that the cell is filled with the solution having, at thermodynamic equilibrium, a bulk density of adsorbable particle $n_0$, and the total number of adsorbable particles remains constant. In certain experimental situation, the sample is connected with a reservoir of the solution to be investigated, and the bulk density of adsorbable particles is maintained at $n_0$. In this case, the condition (\ref{x2}) is no longer necessary since the system is open. In the approximation of constant $n(z)=n_0$ the equation describing adsorption is always (\ref{x5}) with $n_s(t)=n_0$.
Its solution is
\begin{equation}
\label{x5-2}\sigma(t)=\kappa\,\tau\,n_0\left(1-e^{-t/\tau}\right).
\end{equation}
Equation (\ref{x5-2}) indicates that the surface density of adsorbed particles tends to $\kappa \tau n_0$, as in the previous case, with a relaxation time $\tau$, i.e., the relaxation time coincides with the desorption time. The considerations discussed above regarding the limit on $n_0$, expressed by $n_0\ll 1/(\pi R^2 \kappa \tau)$, are also valid in the present case.

\subsection{II.2 With Surface Saturation}
 In the previous subsection, we have considered only the interaction of the adsorbate with the substrate, which we refer to as the direct interaction. In a real problem, however, it is necessary  to take into account the interaction between adsorbate molecules. In other words, the effective adsorption energy results from both the interaction of the adsorbate molecules with the substrate and and their mutual interactions. The interactions between adsorbate molecules can be either attractive or repulsive. In particular, since the adsorbate molecules are not point-like, at sufficiently high surface densities,  hard core repulsion prevents the surface density from exceeding $1/(\pi R^2)$. 
The role of lateral interactions and non-equilibrium effects in surface kinetics has been discussed in \cite{menzel}. The calculation of lateral interactions between adsorbed molecules on planar lattices was presented by Rozebaum \cite{rozebaum}. The coverage dependence of the adsorbate-adsorbate and metal-adsorbate interactions using density functional theory was investigated by Chaudhary et al. \cite{chaudhary}. The results reported in \cite{menzel,rozebaum,chaudhary} indicate that, as expected,  the adsorption depends on the effective adsorption energy and that saturation is eventually reached.
From a phenomenological point of view, this means that the effective adsorption coefficient, $\kappa_{\rm eff}$, in the Langmuir approach depends on the surface density of adsorbed particles. We can therefore write $\kappa_{\rm eff}=\kappa_{\rm eff}(\sigma)$, with the condition that $\kappa_{\rm eff}\to\kappa$ as $\sigma\to 0$.
Denoting by $\sigma_0$ the maximum surface density of adsorbable particles, which is determined by the geometrical dimensions of the adsorbable particles and their lateral interactions \cite{ioannis,liquids}, $\kappa_{\rm eff}$ must vanish as $\sigma\to\sigma_0$. A possible functional form for $\kappa_{\rm eff}(\sigma)$ is then $\kappa_{\rm eff}(\sigma)=\kappa(1-G(\sigma/\sigma_0)$, where $G(0)=0$ and $G(1)=1$.

A first generalization of the model based on (\ref{x4},\ref{x5}) is
\begin{equation}
\label{x14}{\cal A}[n_s(t),\sigma(t)]=\kappa\left(1-\frac{\sigma}{\sigma_0}\right) n_s(t),\quad\quad {\cal D}[n_s(t),\sigma(t)]=\frac{1}{\tau} \sigma(t)
\end{equation}
Expression (\ref{x14}) for the adsorption term is valid when the lateral interactions between adsorbate molecules are repulsive. In this case, as $\sigma$ increases, the effective adsorption coefficient is expected to decrease. Equation (\ref{x1}) can then be written as
\begin{equation}
\label{x15}\frac{d\sigma}{dt}=\kappa\left(1-\frac{\sigma}{\sigma_0}\right)\, n_s(t)-\frac{1}{\tau}\,\sigma(t)
\end{equation}
Note that for $\sigma<\sigma_0$, the effective adsorption coefficient $\kappa(1-\sigma/\sigma_0)>0$, and vice versa for $\sigma>\sigma_0$. In the equilibrium state, assuming again that the bulk density of adsorbable particles is uniform across the sample, the equations defining $n_s$ and $\sigma$ are
\begin{eqnarray}
\label{x16}\kappa\left(1-\frac{\sigma}{\sigma_0}\right)\, n_s-\frac{1}{\tau}\,\sigma&=&0,\\
\label{x17}2\,\sigma+n_s\,d&=&n_0\,d\,\,
\end{eqnarray}
from which it follows
\begin{eqnarray}
\label{x18}n_s&=&\frac{n_0 \kappa \tau-\sigma_0(1+2 \kappa \tau/d)+\sqrt{4 \sigma_0 n_0 \kappa \tau+[\sigma_0(1+2 \kappa \tau/d)-n_0 \kappa \tau]^2}}{2 \kappa \tau},\\
\label{x19}\sigma&=&\frac{\sigma_0(1+2 \kappa \tau/d)+n_0 \kappa \tau-\sqrt{4 \sigma_0 n_0 \kappa \tau+[\sigma_0(1+2 \kappa \tau)-n_0 \kappa \tau]^2}}{4 \kappa \tau/d}.
\end{eqnarray}
In the limit $\sigma_0\to 0$, from (\ref{x18},\ref{x19}) we get $n_s\to n_0-2\sigma_0/d$ and $\sigma\to \sigma_0$
whereas for $\sigma_0\to \infty$, that corresponds to the case considered previously (\ref{x8}).
It is of some interest to investigate the dependence of $n_s$ and $\sigma$ on $n_0$. For $n_0\to 0$, from (\ref{x18},\ref{x19}) we obtain
\begin{equation}
\label{x22}n_s\to n_0\,\frac{1}{1+2(\kappa \tau/d)},\quad\quad \sigma=\frac{n_0 \kappa \tau}{1+2(\kappa \tau/d)},
\end{equation}
showing that in this limit, $n_s$ and $\sigma$ are proportional to $n_0$. In the opposite limit where $n_0\to \infty$, we get
\begin{equation}
\label{x23}n_s\to n_0\left(1-2\,\frac{\sigma_0}{n_0 d}\right),\quad\quad \sigma\to \sigma_0\,\left(1-\frac{\sigma_0}{n_0 \kappa \tau}\right),
\end{equation}
indicating that $n_s\to n_0$ and $\sigma\to \sigma_0$, as expected.

For a sample connected to a reservoir, as discussed above, the equation describing adsorption is obtained from Eq. (\ref{x15}) by substituting $n_s(t)$ with $n_0$, and its solution is given by
\begin{equation}
\label{x15-2}\sigma(t)=\kappa\,T\,n_0\,\left(1-e^{-t/T}\right),
\end{equation}
where the relaxation time $T$ is given by
\begin{equation}
\label{x15-3}T=\frac{\tau}{1+\kappa \tau n_0/\sigma_0}.
\end{equation}
The surface density of adsorbable particles tends to the equilibrium value
\begin{equation}
\label{x15-4}\sigma\to\frac{\kappa \tau n_0}{1+\kappa \tau n_0/\sigma_0}.
\end{equation}
Expression (\ref{x15-4}) has the same form as Langmuir isotherm \cite{x1}.

\subsection{II.3 Generalization}

A simple generalization of the model based on (\ref{x14}) is given by the equations
\begin{equation}
\label{x24}{\cal A}[n_s(t),\sigma(t)]=\kappa\left\{1-\left(\frac{\sigma}{\sigma_0}\right)^\nu\right\} n_s(t),\quad\quad {\cal D}[n_s(t),\sigma(t)]=\frac{1}{\tau} \sigma(t),
\end{equation}
where the exponent $\nu$ could potentially be related to the lateral interactions between the adsorbed molecules. However, no such connection has been established yet. In this case, the kinetic equation takes the form
\begin{equation}
\label{x25}\frac{d\sigma}{dt}=\kappa\left\{1-\left(\frac{\sigma}{\sigma_0}\right)^\nu\right\}\, n_s(t)-\frac{1}{\tau}\,\sigma(t),
\end{equation}
and at equilibrium, the quantities of interest are solutions of the system of equations
\begin{eqnarray}
\label{x26}\kappa\left\{1-\left(\frac{\sigma}{\sigma_0}\right)^\nu\right\}\, n_s-\frac{1}{\tau}\,\sigma&=&0,\\
\label{x27}2\,\sigma+n_s\,d&=&n_0\,d\,
\end{eqnarray}
Solution of (\ref{x26},\ref{x27}) are
\begin{equation}
\label{x28}n_s=\frac{\sigma}{\kappa \tau}\,\frac{1}{1-(\sigma/\sigma_0)^\nu}
\end{equation}
and
\begin{equation}
\label{x29}2 \,\frac{\sigma}{\sigma_0}+\frac{d}{\kappa \tau}\,\frac{\sigma/\sigma_0}{1-(\sigma/\sigma_0)^\nu}=\frac{n_0 d}{\sigma_0}.
\end{equation}
In the case in which the sample is connected to a reservoir, the fundamental equation is still (\ref{x25}) with $n_s(t)=n_0$. At equilibrium the surface density is given by (\ref{x28}) substituting $n_s(t)$ with $n_0$.
\subsection{II.4 Limits and Merits of the Phenomenological Models}
The merit of the models described above is the reduced number of phenomenological parameters ($\kappa$, $\tau$, $\sigma_0$). They work well in the limit of moderate bulk densities of adsorbable particles $n_0$, allowing a continuum description of the medium while remaining consistent with $n_0\ll 1/(\pi R^2 \kappa \tau)$. The predictions of the model are in rather good agreement with the experimental results reported by several researchers relevant to the $\sigma=\sigma(n_0)$, in equilibrium situation and with the time evolution of $\sigma=\sigma(t)$ \cite{x5,x6,x7,x8}.
A strong assumption is that the bulk density of adsorbable particles is uniform across the sample. The description assumes a short-range surface potential responsible for the adsorption forces, which does not enter explicitly in the model. The models are built by "hand" using plausible forms for the adsorption and desorption rates, and  temperature is absent from the description. Some years ago, an attempt was made to connect $\kappa$ and $\tau$ to the surface potential by means of statistical mechanics, leading to simple derived expressions \cite{luiz}.
The goal of the present paper is to investigate the adsorption phenomenon by means of a continuum model. To this end, an expression for the free energy density of the system, taking into account the mixing entropy related to the spatial dependence of the bulk density, is derived. By minimizing the free energy of the system, a profile for the density of particles is obtained, and the surface density of particles is evaluated.

\section{III. Mixing Entropy and Bulk Free Energy Density}
\subsection{III.1 Inhomogeneous Mixing Entropy}
The mixing entropy $S_m$ of a molecular solution is a well known quantity \cite{Cahn58,kittel} for a homogeneous solution. When a wall is introduced in a liquid containing a dispersion of adsorbate particles of bulk density $n_0$ at thermodynamical equilibrium, then in presence of adsorption the bulk density of adsorbate acquires a dependence on the distance $z$ from the wall, that is, $n=n(z)$ and the mixture becomes inhomogeneous. Hereafter, we first calculate the mixing entropy variation of a molecular mixture due to the adsorption. Assuming that the adsorbate particles interact only with the adsorbing substrate with an interaction potential $U(z)$ \cite{israelachvili}, according to statistical mechanics their equilibrium distribution $n(z)$, measured in $\mathrm{m^{-3}}$, is given by
\begin{equation}
\label{1}n=n_0\,e^{-U/k_B T},
\end{equation}
i.e. by the Boltzman distribution. In Eq.(\ref{1}) $k_B$ is the Boltzman constant and  $T$ the absolute temperature. For a given distribution of particles, reversing (\ref{1}) we get the interaction energy
\begin{equation}
\label{2}U=-k_B T \log(n/n_0).
\end{equation}
As it is well known, in the equilibrium state the free energy of the system corresponds to a minimum of the free energy \cite{kittel}, that for the system under consideration it is, per unit volume,
\begin{equation}
\label{3}f=n U- T s,
\end{equation}
where $s$ is the bulk density of entropy. In the equilibrium state from (\ref{3}) we get
\begin{equation}
\label{4}\frac{d f}{dn}=U- T\,\frac{d s}{dn}=0,
\end{equation}
from which it follows
\begin{equation}
\label{5}T\,\frac{d s}{dn}=U.
\end{equation}
Substituting into (\ref{5}) the expression of $U$ given by (\ref{2}) we have
\begin{equation}
\label{6}\frac{d s}{dn}=- k_B \log(n/n_0).
\end{equation}
Consequently the entropy related to the dispersion of the particles is
\begin{equation}
\label{7}s=s_0-k_B n_0\{(n/n_0)\,\log(n/n_0)-(n/n_0)+1\}.
\end{equation} where $s_0$ is the mixing entropy for a homogeneous mixture \cite{kittel}. Equation (\ref{7}) gives the entropy of mixing decrease due to the inhomogeneity introduced in the system by the adsorption.
\subsection{III.2 Bulk free energy density}

The variation of mixing entropy (\ref{7}) was derived assuming that the adsorbate particles interact only with the adsorbing substrate, consequently, the internal energy is entirely described by $U$. In general, however, the internal energy has to account for all relevant interactions. We assume that the additional interactions, not included in $U$, give rise to a bulk energy density, $f_e$, that depends both on the local concentration $n(z)$ and its gradient $\nabla n$, i.e., $f_e=f_e(n,\nabla n)$ \cite{dgb}. For simplicity, we restrict our analysis to a sample in the shape of a slab of thickness $d$, bounded by two identical substrates perpendicular to the $z$-axis at $z=\pm d/2$ in a Cartesian reference frame. In this geometry $f_e=f_e(n,n')$, where $n'=dn/dz$. Within the quadratic approximation
\begin{equation}
\label{8}f_e=f_e(n,n')=\frac{1}{2}\,k\,n'^2+\frac{1}{2}\beta (n-n_0)^2,
\end{equation}
where $k$ and $\beta$ are two phenomenological parameters, measured in J$\cdot$ m$^5$ and J$\cdot$ m$^3$, respectively. The first contribution, related to $n'^2$, arises from the assumption that the uniform state is stable \cite{landau}, while the second contribution, proportional to $(n-n_0)^2$, reflects the assumption that, in the absence of adsorption, the equilibrium bulk density of adsorbate particles is $n_0$. The derivation of (\ref{8}) is presented in Appendix 1.

In the absence of adsorption, $n(z)=n_0$ across the sample for $-d/2\leq z \leq d/2$. In the presence of adsorption, $n=n(z)$ while the conservation of the total number of particles imposes the condition
\begin{equation}
\label{9}\int_{-d/2}^{d/2} n(z)\,dz=n_0 d
\end{equation}
\section{IV. The Variational Problem }
\subsection{IV.1 Total Free Energy of the Sample}
In the state of thermodynamical equilibrium, the total free energy, $F$, of the system, which includes the bulk energy density $f(n,n')=f_e(n,n')+n U$ and the entropic contribution $-T\,s(n)$ reaches a minimum value \cite{dgb}. This problem has also been discussed in the context of ferro-nematics by Burylov and Raikher et al.\cite{raiker}. $F$ is given by
\begin{equation}
\label{10}F=\int_{-d/2}^{d/2} \left\{f(n,n')-T\,s(n)\right\} dz.
\end{equation}
Due to the constraint (\ref{9}), the quantity to be minimized is
\begin{equation}
\label{11}G=\int_{-d/2}^{d/2} g(n,n')\, dz,
\end{equation}
where
\begin{equation}
\label{12}g(n,n')=f(n,n')-T\,s(n)+\, \lambda n,
\end{equation}
where $\lambda$ is a Lagrange multiplier associated with the constraint (\ref{9}) \cite{elsgots}. According to the variational calculus, the function that minimizes (\ref{11}) is a solution of Euler-Lagrange equation
\begin{equation}
\label{13}\frac{d}{dz}\,\frac{\partial g}{\partial n'}-\frac{\partial g}{\partial n}=0,
\end{equation}
with boundary conditions
\begin{equation}
\label{14}\frac{\partial g}{\partial n'}=0,\quad{\rm for}\quad z=\pm d/2,
\end{equation}
corresponding to the free borders \cite{elsgots}.
The simple case in which the sample occupies the region $z\geq 0$ corresponds to the half-space approximation. The detailed analysis is presented in Appendix 2. In this situation, using a Cartesian reference frame with the origin at the limiting surface, the boundary conditions of the problem are
\begin{equation}
\label{14-1}\frac{\partial g}{\partial n'}=0,\quad{\rm for}\quad z=0,
\end{equation}
and
\begin{equation}
\label{14-2}\lim_{z\to \infty}n(z)=n_0,
\end{equation}
since very far from the adsorbing surface, the liquid is in its equilibrium state. In this framework, the condition (\ref{9}) is no longer necessary.

\subsection{IV.2 Solution of the Linearized Problem for the Slab}

Taking into account that  $f(n,n')=f_e(n,n')+n U$, and (\ref{8}), Eq.(\ref{13}) gives
\begin{equation}
\label{15}k n''-\{\beta (n-n_0)+U+k_B T \log(n/n_0)+\lambda\}=0,
\end{equation}
to be solved with the boundary conditions (\ref{14}), that for the present problem are
\begin{equation}
\label{16} n'(\pm d/2)=0.
\end{equation}
In the following, we assume that the interaction energy of the adsorbable particles with the substrates is of the type
\begin{equation}
\label{17}U=V_0 \frac{\cosh(z/L)}{\cosh(d/2L)},
\end{equation}
where $L$ is the range of the surface forces \cite{israelachvili}. Furthermore, we restrict our analysis to the case in which the adsorption induces only a small deviation of the bulk particle density from $n_0$. Defining $x=n/n_0$, and writing $x=1+\epsilon$, where $\epsilon$ represents the relative variation of the particle density due to adsorption, we restrict our analysis to the case $|\epsilon|\ll 1$.
In this framework, Eq.(\ref{15}) can be rewritten as
\begin{equation}
\label{18}k n_0 \epsilon''-\{\omega\,\epsilon +U(z)+\lambda\}=0,
\end{equation} where $\omega = k_B T + n_0\,\beta$,
to be solved with the constraint
\begin{equation}
\label{19}\int_{-d/2}^{d/2}\epsilon(z)\,dz=0
\end{equation}
and the boundary conditions
\begin{equation}
\label{20}\epsilon'(\pm d/2)=0.
\end{equation}
Inspection of (\ref{18}) shows that the compressibility contribution associated with $\beta$ becomes comparable to the entropic contribution when
$\beta\sim k_B T/n_0$.
It follows that for a given $n_0$, one can define an intrinsic compressibility coefficient defined $\beta_s= k_B T/n_0$,
such that $\beta<\beta_s$ and $\beta>\beta_s$ correspond to small or large compressibility, respectively. Furthermore, taking into account the form assumed for $U=U(z)$, the compressibility term becomes comparable with the adsorption term when
$\beta\sim V_0/n_0$,
allowing us to define another critical value for the compressibility coefficient
$\beta_a=V_0/n_0$.
Finally, the entropic and adsorption terms are comparable when
$V_0\sim k_B T$.

At this stage, it is convenient to define a characteristic length scale associated with the problem:
\begin{equation}
\label{22}\xi=\sqrt{k/\beta_{eff}},
\end{equation}
where $\beta_{eff}=\omega/n_0$ is an effective compressibility which includes an entropic contribution. $\xi$ is a correlation length related to the elastic and compressibility properties of the system, indicating the penetration length for a density variation produced in the bulk \cite{durand}.

The solution of the ordinary differential Eq.(\ref{18}), where $U(z)$ is given by (\ref{17}), and satisfying the boundary conditions (\ref{19}) and (\ref{20}) is
\begin{equation}
\label{23}\epsilon=\frac{L V_0}{d(L^2-\xi^2)\omega}\,\left\{- d L\frac{\cosh(z/L)}{\cosh(d/2L)}+\left(2( L^2- \xi^2)+d \xi\,\frac{\cosh(z/\xi)}{\cosh(d/2\xi)}\right)\tanh(d/2L)\right\},
\end{equation}
while the Lagrange's multiplier is found to be
\begin{equation}
\label{22}\lambda=-2\,\frac{L}{d}\,V_0\,\tanh(d/2 L),
\end{equation}
From (\ref{22}), it follows that in the half space limit, where $d\to \infty$, $\lambda=0$.

\begin{figure}[!htb]
    \centering
    \includegraphics[width=.6\textwidth]{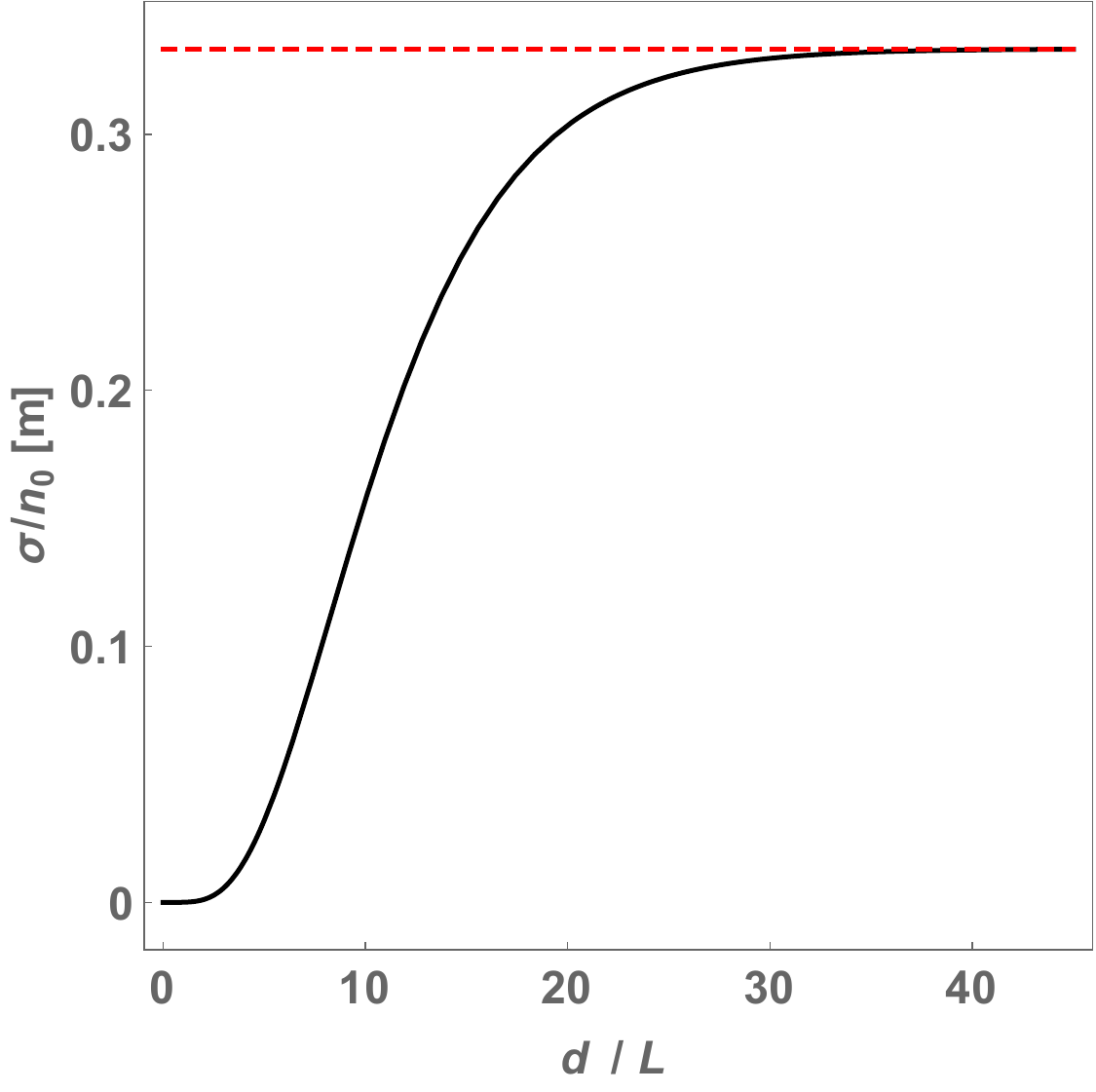}
    \caption{Thickness dependence of $\sigma$, defined as the fist moment of $\epsilon(z)$, versus the thickness of the sample. The horizontal dashed line corresponds to the value given by (\ref{25}) for $d\to \infty$. The curve is drawn for $V_0/\omega=-1/3$ and $\xi=2\, L$.
    }
    \label{figure_1}
\end{figure}

In the case of interest, $d\gg 2L$ and $\xi\sim L$. Consequently, $\epsilon=\epsilon(z)$ is nonzero only within two surface layers whose thickness is of the order of $L$. This means that $n(z)$ differs from $n_0$ only within the surface layers, making it possible to introduce a surface density of adsorbed particles defined by
\begin{equation}
\label{24}\sigma=\int_0^{d/2}[n(z)-n(0)]dz=n_0\int_{0}^{d/2}[\epsilon(z)-\epsilon(0)]dz,
\end{equation}
where $n(0)$ is the bulk density of particles at the middle of the sample. Using (\ref{24}) together with (\ref{23}), we get
\begin{equation}
\label{25}\sigma=-\frac{L n_0 V_0}{2(L^2-\xi^2)\omega}\,\left\{-\frac{d L}{\cosh(d/2L)}+\left(2(L^2- \xi^2)+\frac{d \xi}{\cosh(d/2\xi)}\right)\,\tanh(d/2 L)\right\}
\end{equation}
From (\ref{25}) it follows that, for an attractive potential, i.e. $V_0<0$, $\sigma>0$, indicating that the surface adsorbs particles, whereas for a repulsive potential $\sigma<0$. In Figure \ref{figure_1}, we show the thickness dependence of $\sigma$ given by (\ref{25}). It is important to note that
$\sigma$ already reaches its limiting value for $d>30 L$.

We stress that defining  $\sigma$ as the first moment of
$\epsilon$ appears reasonable, although this definition is not unique. The issue of defining  $\sigma$ will be discussed in Appendix 2, since it strongly depends on the experimental technique used to detect variations in the bulk density of the adsorbable particles.
\subsection{IV.3 Half Space Approximation}
In the half space approximation, $d\to \infty$, and from (\ref{25}) and (\ref{22}) we obtain
\begin{equation}
\label{26}\sigma_{\rm hs}=-\frac{L V_0}{\beta_{eff}}.
\end{equation}
This result means that the surface density of adsorbed particles is independent of the elastic constant $k$. Nevertheless, one should note that this unexpected result is a consequence of the linear approximation. If higher order terms are retained in the approximation, additional $k$ dependent terms appear, contributing corrections to the first order approximation given above. It is of some interest to compare (\ref{26}) with (\ref{x15-4}). From the comparison, it follows that
\begin{equation}
\label{f}\kappa \tau=-\,L\,\frac{V0}{k_{B}T},\quad{\rm and}\quad\sigma_0=-L\,\frac{V_0}{\beta},
\end{equation}
indicating that the adsorption length $\kappa \tau$ is proportional to the range of the surface forces responsible for the adsorption, and that the maximum surface density depends on the compressibility, as expected.

Finally, it is interesting to observe that, for
\begin{equation}
\label{27}n_0\to 0, \quad \sigma_{\rm hs}=-\frac{n_0 L V_0}{k_B T},
\end{equation}
the surface density of adsorbed particles is proportional to the bulk density of adsorbate particles in thermodynamic equilibrium. On the other hand, for
\begin{equation}
\label{28}n_0\to \infty, \quad \sigma_{\rm hs}=-\frac{L V_0}{\beta},
\end{equation}
indicating that $\sigma$ tends to a saturation value \cite{langmuir}. An analysis of the half-space approximation is provided in Appendix 2

\subsection{IV.4 Localized surface energy}

\begin{figure}[!htb]
    \centering
    \includegraphics[width=.6\textwidth]{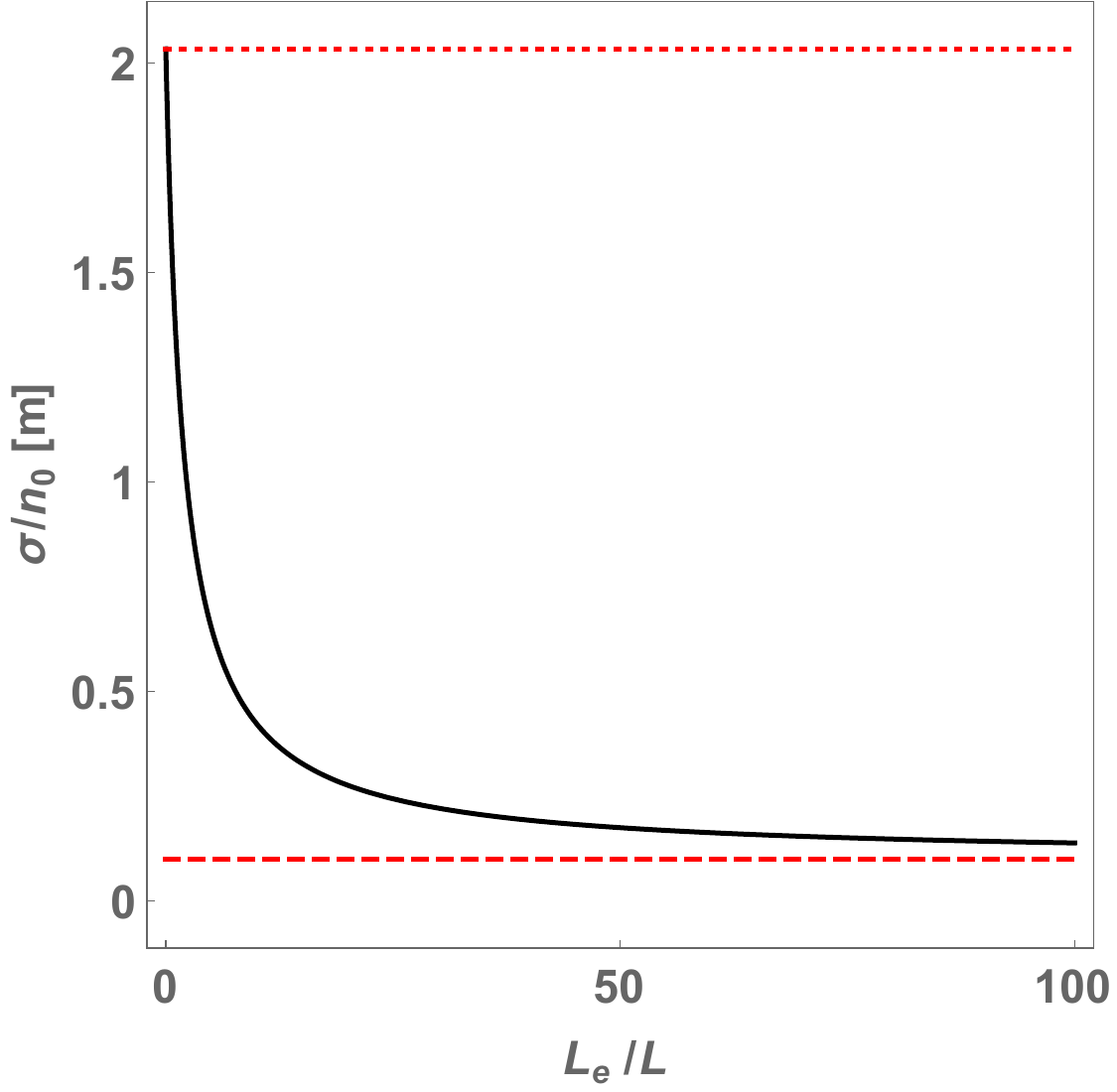}
    \caption{Dependence of $\sigma$, defined as the fist moment of $\epsilon(z)$, versus the extrapolation length of the localized surface energy $L_e$. The horizontal dashed line corresponds to the case where $L_e\to \infty$, i.e. to the absence of localized surface energy. In this case $\sigma$ is given by (\ref{25}). The dotted line corresponds to the case of strong surface localized energy where $L_e\to 0$, and the surface density is given by (\ref{351}). The curve is drawn for $V_0/\omega=-1/3$, $n_e=2 n_0$, and $\xi=2\, L$.
    }
    \label{figure_2}
\end{figure}

In the analysis presented above, the substrate is responsible for the interaction $U(z)$, which we have assumed to have the functional form (\ref{17}), characterized by the amplitude $V_0$ and the penetration length $L$. The interaction energy $U(z)$ is localized within a surface layer of thickness $L$, which we have taken to be of the order of $\xi$. In some cases, it is also necessary to take into account a localized surface energy related, for instance, to hard-core repulsion\cite{raikher1}. The corresponding surface energy, $f_s$, exhibits a minimum for a well defined value of the bulk density at the surface, denoted by $n_e$. Within our framework, where the two confining substrates are identical, $f_s$ and $n_e$ are the same for both limiting surfaces. A possible form for $f_s$, in the quadratic approximation, is given by
\begin{equation}
\label{29}f_s=f_0(n_e)+\frac{1}{2}\,w\,(n_s-n_e)^2,
\end{equation}
where $f_0$ is the surface energy in the equilibrium state, $w$  is the surface energy strength, measured in $\mathrm{J \cdot m^4}$, and $n_s=n(\pm d/2)$. In the present case the equilibrium state corresponds to the minimum of the quantity
\begin{equation}
\label{30}G=\int_{-d/2}^{d/2} g(n,n')\, dz+\frac{1}{2}\,w\,[n_s(-d/2)-n_e]^2+\frac{1}{2}\,w\,[n_s(d/2)-n_e]^2,
\end{equation}
instead of (\ref{11}). The Euler-Lagrange equation relevant to (\ref{30}) is again given by (\ref{15}), to be solved with the boundary conditions \cite{elsgots}
\begin{equation}
\label{31}\pm\,\frac{\partial g}{\partial n'}+\frac{\partial f_s}{\partial n_s}=0,\quad{\rm for}\quad z=\pm d/2.
\end{equation}
Taking into account (\ref{29}), (\ref{8}) and (\ref{12}), Eqs. (\ref{31}) can be rewritten as
\begin{equation}
\label{32}\pm\,k\, n'+\,w\,(n_s-n_e)=0,\quad{\rm for}\quad z=\pm d/2.
\end{equation}
Introducing $\epsilon$, (\ref{32}) takes the form
\begin{equation}
\label{33}\pm\,\mu\,\epsilon'(\pm d/2)+\,n_0\,\epsilon(\pm d/2)=n_0-n_e,
\end{equation}
where $\mu=L_e n_0$, and $L_e=k/w$ is the surface extrapolation length \cite{durand}.
The problem can be solved following the approach  described in the preceding sections. In the following, we restrict the analysis to the surface density of adsorbed particles in the half-space limit, $d\to \infty$. In this limit we obtain
\begin{equation}
\label{35}\sigma_{\rm hs}=\frac{-L n_0 V_0[L \xi+L_e(L+\xi)]+(n_e-n_0)\,\xi^2(L+\xi) \omega}{(L+\xi)(L_e+\xi)\omega},
\end{equation}
which for $w=0$, i.e. $L_e\to \infty$, reduces to (\ref{26}), as expected. The dependence of $\sigma$ given by (\ref{35}) on $L_e$ is shown in Figure \ref{figure_2}.
In the case $w\to \infty$, which implies $L_e\to 0$, from (\ref{35}) we get
\begin{equation}
\label{351}\sigma_{\rm hs}=\frac{\omega \xi^2(n_e-n_0)-L[L n_0 V_0-(n_e-n_0)\xi \omega]}{(L+\xi)\,\omega}.
\end{equation}

In the limit $V_0=0$, i.e. in the case where there is no delocalized surface energy, it follows from (\ref{27}) that
\begin{equation}
\label{28}\sigma_{\rm hs}(V_0=0)=\frac{(n_e-n_0) \xi^2}{L_e+ \xi}.
\end{equation}
Equation (\ref{28}) indicates that a localized surface energy could be responsible for an accumulation of particles at the surface, even in the absence of an adsorption energy $U$.  In this situation, it is possible to define an equivalent adsorption parameter $(L V_0)_{\rm eq}$ by comparing (\ref{28}) with (\ref{26}). A simple calculation gives
\begin{equation}
\label{29} (L V_0)_{\rm eq}=-\,\frac{\omega\,\xi^2}{n_0(L_e+\xi)}\,\,(n_e-n_0).
\end{equation}
For $L_e\to \infty$, $(L V_0)_{\rm eq}\to 0$, as expected. In the opposite limit $w\to \infty$, i.e. $L_e\to 0$, corresponding to a very large surface energy strength,
\begin{equation}
\label{30} (L V_0)_{\rm eq}=-\,\frac{\omega \xi}{n_0}\,(n_e-n_0).
\end{equation}
Adsorption occurs only for $n_e>n_0$, as expected.
\section{V. Conclusion}
Adsorption at a solid–liquid interface has been considered. We have assumed the presence of an adsorption energy localized within a surface layer of mesoscopic thickness. The bulk free energy density of the liquid, in which the ions are dissolved, has been described within the quadratic approximation, consisting of an elastic term proportional to the square of the density gradient and a term quadratic in the deviation from the thermodynamic equilibrium density.
The equilibrium density distribution of adsorbate particles has been obtained by minimizing the free energy of the system, under the constraint of particle conservation, which introduces a Lagrange multiplier, as well as the mixing entropy. The surface density of adsorbed particles has then been derived from the equilibrium bulk density profile, assuming that the forces responsible for adsorption are short-ranged. The case including a localized surface energy has also been considered.
To obtain an analytical solution, we have restricted the analysis to the regime in which adsorption induces only small deviations of the bulk density from its equilibrium value, allowing linearization of the Euler–Lagrange equation. In the general case of strong adsorption one has to considered the non-linear Euler–Lagrange equation which has to be  solved numerically.

\section{Appendix 1}
The purpose of this Appendix is to justify Eq.(\ref{8}). Let us denote by $u(r)$ the intermolecular interaction potential between two adsorbate molecules at a distance $r$, and assume that $u(r)$ has a minimum at $r=r_0$. This implies that, at equilibrium, the bulk density is given by $n_0 = 1/r_0^3$ in the case of a cubic lattice. Under these assumptions,
\begin{equation}
\label{a1}u(r_0)={\rm minimum}\,\to \left(\frac{du}{dr}\right)_{r_0}=0,\quad{\rm and}\quad \left(\frac{d^2 u}{dr^2}\right)_{r_0}=H>0
\end{equation}
If, through a homogeneous compression or dilatation, the distance between two interacting molecules changes from $r_0$ to $r=r_0+\Delta r$, the intermolecular interaction changes from $u(r_0)$ to $u(r)$. Assuming $|\Delta r| \ll r_0$, i.e. a small compression, we can expand $u(r)$ as
\begin{equation}
\label{a2}u(r)=u(r_0)+\left(\frac{du}{dr}\right)_{r_0}\,\Delta r+\frac{1}{2}\left(\frac{d^2 u}{dr^2}\right)_{r_0}\,(\Delta r)^2,
\end{equation}
which, using Eq. (\ref{a1}), can be rewritten in the form
\begin{equation}
\label{a3}u(r)=u(r_0)+\frac{1}{2}\,H\,(\Delta r)^2.
\end{equation}
Denoting by $\gamma$ the number of first neighbors ($\gamma=6$ for a simple cubic lattice), the total energy per molecule is
${\cal U}(r_0)=\gamma u(r_0)/2$,
in the ground state, and
${\cal U}(r)=\gamma u(r)/2$,
in the compressed state. It follows that the excess energy per molecule due to the compression is
\begin{equation}
\label{a6}\Delta {\cal U}={\cal U}(r)-{\cal U}(r_0)=\frac{1}{4}\,\gamma\,H\,(\Delta r)^2.
\end{equation}
Taking into account that
$n_0=1/r_0^3$, $n=1/r^3$,
we obtain
$r_0=n_0^{-1/3}$,  $r=n^{-1/3}$, and consequently
\begin{equation}
\label{a9}\Delta r=r-r_0=\frac{r_0}{3}\,\,\frac{n_0-n}{n_0}.
\end{equation}
It follows that
\begin{equation}
\label{a10}\Delta{\cal U}=\frac{1}{4}\,\gamma\,H\,\left(\frac{r_0}{3}\right)^2\,\left(\frac{n_0-n}{n_0}\right)^2.
\end{equation}
The energy density, $h$, is simply given by $h=n {\cal U}$, while the excess energy density due to the compression is $f=n \,\Delta{\cal U}(r)$, which, within the quadratic approximation, is proportional to $(n-n_0)^2$, as assumed in (\ref{8}).

Let us now consider the case in which the bulk density of adsorbable particles depends on position. As before, we restrict our analysis to a one-dimensional problem, where $n=n(z)$. We denote by $z$ and $z+\Delta z$ the coordinates of two interacting particles, and by $n(z)$ and $n(z+\Delta z)$ the corresponding bulk densities. In our approximation, $\Delta z \sim r_0$, since in the absence of spatial dependence of the particle density, the distance between two interacting particles is $r_0$. Within this framework, $u(r)=u(n(z))$.
Assuming a slow variation of $n(z)$, we write $n(z)=n_0+\Delta n(z)$, where $\Delta n(z)=r_0\,\,n'(z) $. It follows that the distance between the interacting molecules varies with $z$, and is given by
\begin{equation}
\label{a12}\Delta r(z+r_0)=n^{-1/3}(z+r_0)-n^{-1/3}(z)=-\frac{r_0^2}{3 n_0}\,n'(z)
\end{equation}
In this case,
\begin{equation}
\label{a13}\Delta{\cal U}=\frac{1}{4}\,\gamma\,H\,\left(\frac{r_0^2}{3 n_0}\right)^2\,n'(z)^2,
\end{equation}
indicating that the bulk variation of the interaction energy density is proportional to $n'^2(z)$, as assumed in deriving (\ref{8}).
\section{Appendix 2}
\subsection{semi-infinite sample}

The analysis presented in the main text makes it possible to investigate the dependence of $\sigma$ on the sample thickness $d$. This dependence may be important since, from $\sigma=\sigma(d)$, it is possible to derive the parameters characterizing the adsorption phenomenon. For completeness, in this Appendix we consider the case of a semi-infinite sample, for which the mathematical analysis is simpler and some conclusions can be drawn regarding the role of $k$, $\beta$, $L$, and $V_0$ in determining $\sigma$.

Let us consider the case in which the sample occupies the region $0 \leq z < \infty$ in a Cartesian coordinate system with the adsorbing surface located at $z=0$. We assume that the interaction energy between the adsorbate particles and the adsorbant substrate has the form
\begin{equation}
\label{b1}
U(z)=V_0\,e^{-z/L},
\end{equation}
with amplitude $V_0$ and penetration length $L$, which is an extention of Eq. (\ref{17}) to the case of a semi-infinite sample.
As before, we assume that the perturbation of the bulk density of the adsorbate particles induced by adsorption is small. Hence, we set $n/n_0 = 1 + \epsilon$, with $\epsilon \ll 1$. Within this framework, $\epsilon(z)$ satisfies the ordinary differential equation
\begin{equation}
\label{b2} k n_0 \epsilon''(z)-[\omega\,\epsilon(z)+V_0\,e^{-z/L}]=0,
\end{equation} where the Lagrange multiplier $\lambda$ is absent since we have relaxed the constraint on the conservation of the number of particles in Eq. (\ref{9}). The boundary conditions for Eq. (\ref{b2}) are
\begin{equation}
\label{b3}\epsilon'(0)=0,\quad{\rm and}\quad \lim_{z\to \infty}\epsilon(z)=0.
\end{equation}
We rewrite Eq.(\ref{b2}) as
\begin{equation}
\label{b4}\epsilon''(z)-\left\{\frac{1}{\xi^2}\,\epsilon(z)+\frac{V_0}{k n_0}\,e^{-z/L}\right\}=0,
\end{equation}
where $\xi$ is defined in (\ref{22}). The solution of (\ref{b4}) with the boundary conditions (\ref{b3}) is
\begin{equation}
\label{b5}\epsilon(z)=\,\frac{\xi^2 L}{\xi^2-L^2}\,\frac{V_0}{k n_0}\,\left(L\,e^{-z/L}-\xi\,e^{-z/\xi}\right).
\end{equation}
The above expression shows that the profile $\epsilon(z)$ is the superposition of two exponentials: one related with the correlation length ($\xi$) of the bulk and another related with the range of the surface forces ($L$).
\begin{figure}[!htb]
    \centering
    \includegraphics[width=.6\textwidth]{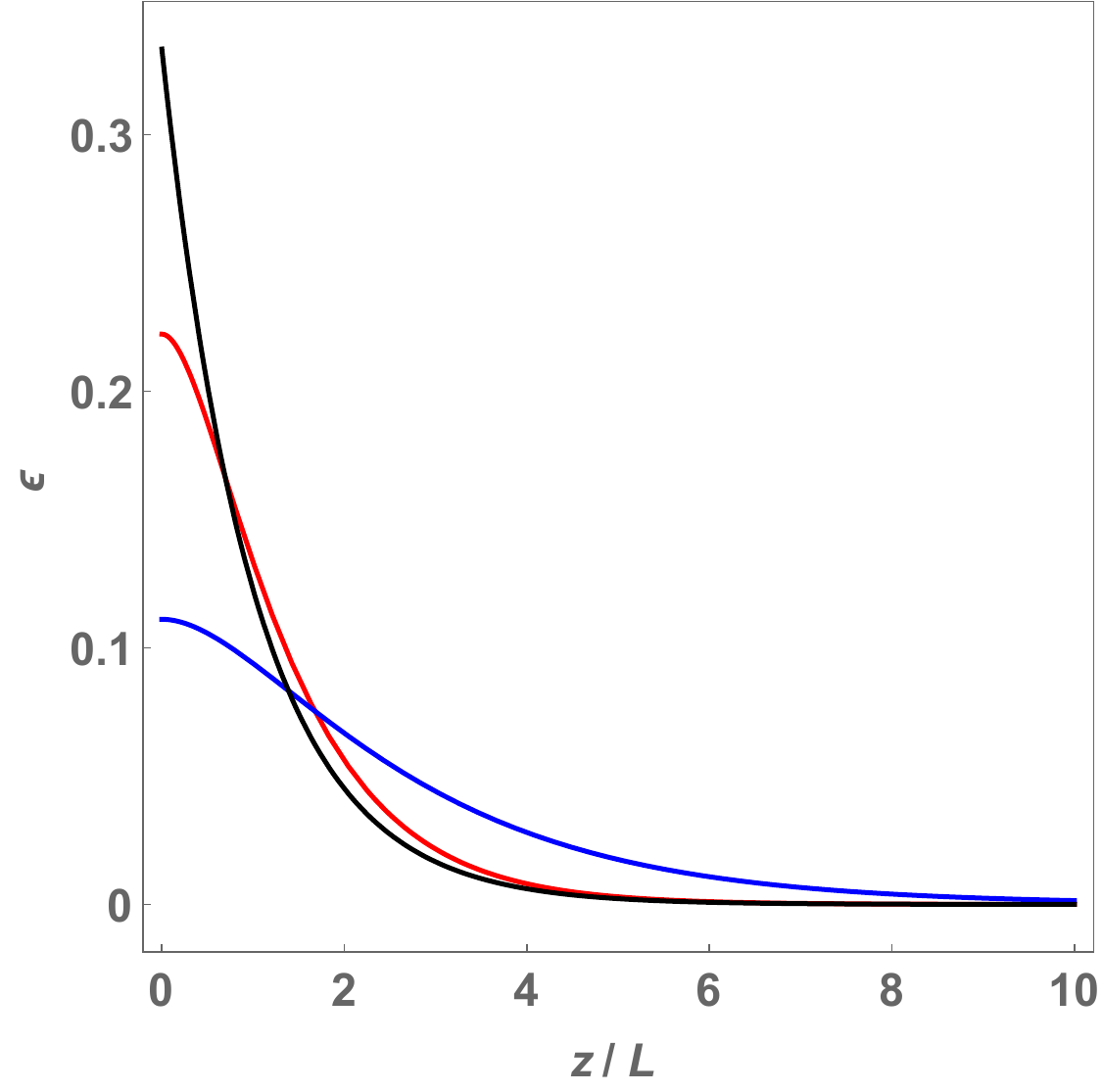}
    \caption{Influence of the elastic constant $k$ on $\epsilon=n/n_0-1$. Black curve $\xi=0$, red curve $\xi=0.5\times L$, blue curve $\xi=2\times L$. The curve are drawn for $V_0/\omega=-1/3$.
    }
    \label{figure_3}
\end{figure}
In the limit $\xi\to L$, we get
\begin{equation}
\label{b6}\epsilon(z)=-\frac{L}{2}\,\frac{V_0}{k n_0}\,(z+L)\,e^{-z/L}.
\end{equation}

\subsection{dependence of $\epsilon$ on the elastic constant}
To investigate the dependence of $\epsilon=\epsilon(z)$ on the elastic constant $k$, we rewrite $k$ in terms of $\xi$. From the definition of $\xi$, we obtain $\xi^2 = k n_0/\omega$, and
 (\ref{b5}) can be rewritten as
\begin{equation}
\label{b62}\epsilon(z)=\frac{V_0L}{\omega}\,\frac{L\,e^{-z/L}-\xi\,e^{-z/\xi}}{\xi^2-L^2}.
\end{equation}
For $z=0$, from (\ref{b62}) we obtain
\begin{equation}
\label{b63}\epsilon(0,\xi)=-\frac{V_0}{\omega}\,\frac{L}{\xi+L}.
\end{equation}
From (\ref{b63}), it follows that the surface value of $\epsilon$ decreases as the elastic constant increases.

We observe that in the case of $k=0$, that implies $\xi=0$, from (\ref{b2}), or from (\ref{b63}), we get
\begin{equation}
\label{b64}\epsilon(z,\xi=0)=-\frac{V_0}{\omega}\,e^{-z/L},
\end{equation}
and hence $\epsilon'(0,\xi=0)=V_0/(\omega L)\neq 0$.
We note that in the case $k=0$ the boundary condition (\ref{14}) is automatically verified since it is actually $k\,n_0\,\epsilon'(0)=0$.
In Figure \ref{figure_3} we show the $\epsilon=\epsilon(z)$ for three values of the elastic constant $k$. As expected, increasing the elastic constant decreases the surface value of $\epsilon$, according to (\ref{b63}), and the "deformation" propagates into the bulk over a longer distance.


The equivalent surface density of adsorbed particles is given by
\begin{equation}
\label{b7}\sigma=n_0 \int_0^{\infty}\,\epsilon(z) dz=-L V_0\,\frac{n_0}{\omega}
\end{equation}
which coincides with Eq.~(\ref{26}). It is important to emphasize that $\sigma$, within the linear approximation, is independent of $\xi$, i.e., of the elastic constant $k$. This parameter only sets the characteristic length scale over which the bulk density of adsorbable particles is perturbed.

An alternative definition of $\sigma$ can be introduced via the second moment of $\epsilon(z)$,
\begin{equation}
\label{b71}\delta=\int_0^{\infty} \epsilon^2(z)\,dz,
\end{equation}
which now depends on $\xi$, as a simple calculation, using for $\epsilon(z)$ expression (\ref{b62}), yields
\begin{equation}
\label{b72}\delta=\frac{V_0^2\,L^2}{3\,\omega^2}\,\frac{L^2+3 L \xi+\xi^2}{(L+\xi)^3}.
\end{equation}
This suggests that if an equivalent surface density, $\sigma_{2}$, is defined as $\sigma_{2}\propto \sqrt{\delta}$, then
\begin{equation}
\label{b73}\sigma_{2}\propto\frac{V_0 L}{\omega (L+\xi)}\,\,\sqrt{\frac{L^2+3 L \xi+\xi^2}{3 (L+\xi)}}.
\end{equation}
\begin{figure}[!htb]
    \centering
    \includegraphics[width=.6\textwidth]{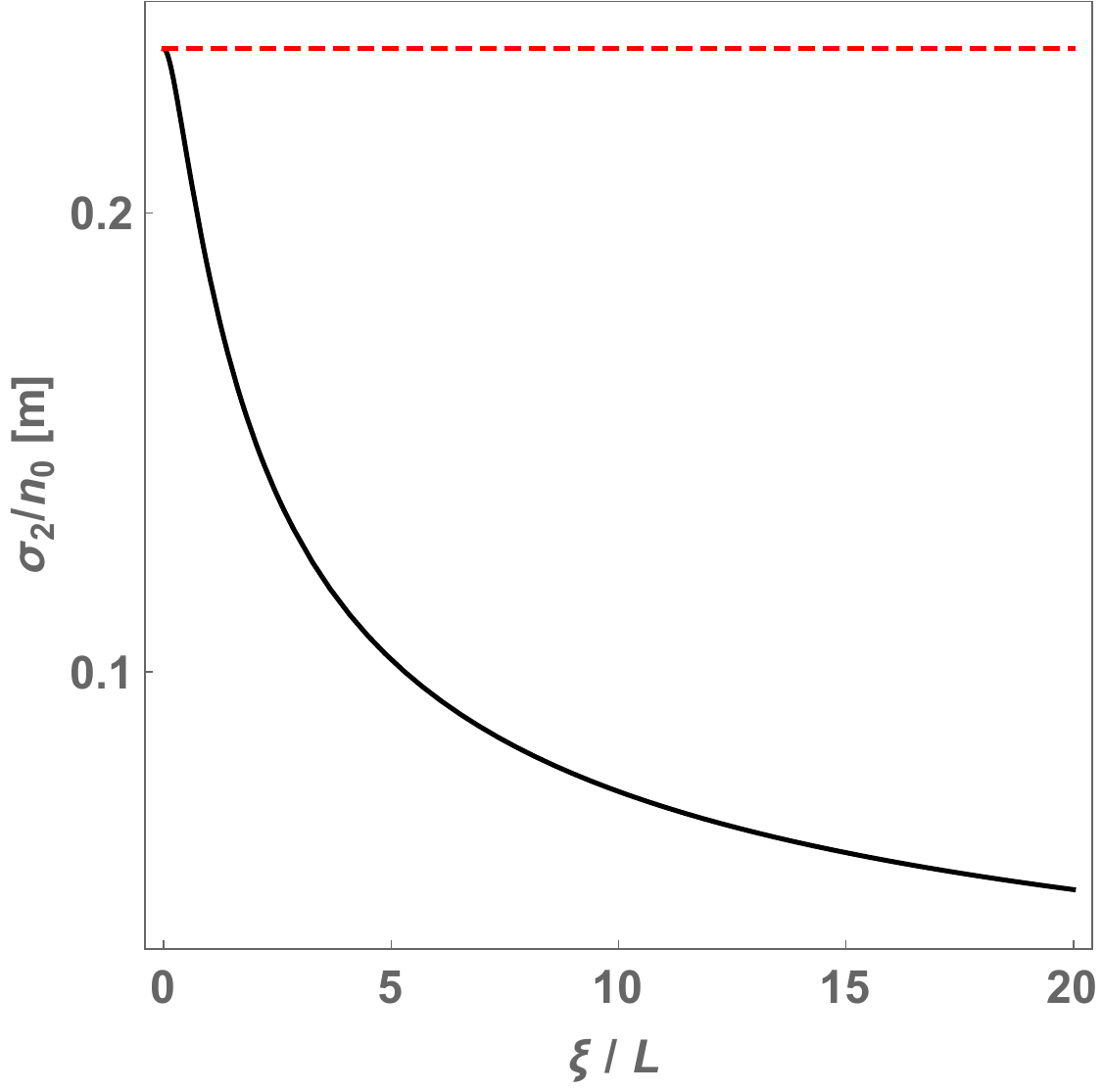}
    \caption{Dependence of $\sigma_{2}\propto \sqrt{\delta}$ on $\xi$. The curve are drawn for  $V_0/\omega=-1/3$. The horizontal line corresponds to $k=0$.}
    \label{figure_4}
\end{figure}
In Figure \ref{figure_4} we show $\sigma_2=\sqrt{\delta}$ versus $\xi$ (red line). The case in which $k=0$, corresponding to $\xi=0$ is also reported (blue line). The above two definitions of $\sigma$ correspond to different moments of the adsorbed particle distribution. The first definition informs us about the total number/mass of adsorbed particles at a surface, but it provides no information about how these particles are spatially distributed. The second definition, $\sigma_2$ contains structural information, as it depends on the distribution profile of the adsorbed particles. An example is illustrated in Figure 3, where although the area under each of the three curves, corresponding to the total number of adsorbed particles, is the same, the shape of the profiles differs, resulting in different values of the higher-order moment $\sigma_2$. Experimentally, the total mass of the adsorbed particles can be measured by techniques as the quartz crystal microbalance, optical reflectometry, surface plasmon resonance, etc, while the mass distribution, the profil, is measured by variable-angle total internal reflexion fluorescence, neutron and X-ray specular  scattering techniques \cite{neutronR,Russell}.

\subsection{localized surface energy}

Finally, in the presence of localized surface energy of the type $f_s=(w/2)(n_s-n_e)^2$, where $n_s=n(0)$, Eq.(\ref{b2}) has to be solved with the boundary conditions
\begin{equation}
\label{b8}-n_0\,L_e\,\epsilon'(0)+n_0\,\epsilon(0)+(n_0-n_e)=0,\quad{\rm and}\quad \lim_{z\to \infty}\epsilon(z)=0.
\end{equation}
 The solution is
\begin{equation}
\label{b9}\epsilon(z)=\epsilon_1(z)+\epsilon_2(z),
\end{equation}
where
\begin{eqnarray}
\label{b10}\epsilon_1(z)&=&-\,\frac{L^2 \xi^2 V_0}{k n_0 (L^2-\xi^2)}\,e^{-z/L},\\
\label{b11}\epsilon_2(z)&=&\xi\,\frac{L(L+L_e) \xi^2 V_0-k(n_0-n_e)(L^2-\xi^2)}{k n_0 (L_e+\xi)(L^2-\xi^2)}\,e^{-z/\xi}.
\end{eqnarray}
In this case the equivalent surface density of adsorbed particles is
\begin{equation}
\label{b12}\sigma=-\xi^2\,\frac{k(n_0-n_e)(L+\xi)+L V_0 [L_e\xi+L(L_e+\xi)]}{k (L+\xi)(L_e+\xi)},
\end{equation}
that for $w=0$, i.e. for $L_e\to \infty$, reduces to (\ref{b7}).

Finally, for $V_0=0$ from (\ref{b12}) we get
\begin{equation}
\label{b13}\sigma=-\xi^2\,\frac{n_0-n_e}{L_e+\xi}.
\end{equation}
It follows that a localized surface energy is equivalent to an adsorption energy
\begin{equation}
\label{b14}(L V_0)_{\rm eq}=\frac{\omega}{n_0}\,\xi^2 \frac{n_0-n_e}{L_e+\xi},
\end{equation}
as it follows comparing (\ref{b13}) with (\ref{b7}).


\end{document}